\documentclass{epl}
\usepackage{amsmath}
\renewcommand{\epsilon}{\varepsilon}
\renewcommand{\phi}{\varphi}

\newcommand{\xieq}{\xi_{\rm eq}}

\title{Surfing on a critical line:
Rejuvenation without chaos,
Memory without a hierarchical phase space}
\author{Ludovic Berthier 
\and Peter C. W. Holdsworth}
\institute{
Laboratoire de Physique, ENS-Lyon and CNRS, 
46 All\'ee d'Italie,
69007 Lyon, France
}

\begin{document}

\maketitle

\begin{abstract}
The dynamic behaviour of glassy materials displays strong
nonequilibrium effects, such as ageing in simple protocols,
memory, rejuvenation and Kovacs effects in 
more elaborated experiments.
We show that this phenomenology may be easily understood
in the context of the nonequilibrium critical dynamics
of non-disordered systems,
the main ingredient  being the existence 
of an infinite equilibrium correlation length.
As an example, we analytically investigate 
the behaviour of the 2D XY model submitted to temperature
protocols similar to experiments.
This shows that typical glassy effects may be
obtained by `surfing on a critical line' without
invoking the concept of temperature chaos nor 
the existence of a hierarchical phase space,
as opposed to previous theoretical approaches.
The relevance of this phenomenological approach
to glassy dynamics is finally discussed.
\end{abstract}

\vspace*{-1.cm}
{\it ...comme des amateurs de surf qui glissaient sur la vague.}
Vassili Axionov, {\it Une br\^ulure}.

\vspace*{.5cm}

Experiments on a large class of glassy materials demonstrate
intriguing similarities in their nonequilibrium dynamic
properties.
Structural ---in particular polymeric--- glasses are
the paradigm, and have been studied in a systematic way 
over a long period time~\cite{struik}. 
Interest in ageing experiments grew 
with the realization that
very similar effects are observed in
spin glasses~\cite{SG}.
More recently, the family of glassy systems has been much enlarged
to include systems as different as
dirty type-II superconductors~\cite{supra}, several complex 
fluids~\cite{complex}, disordered dielectrics~\cite{diel,relax}
and ferromagnets~\cite{disferro},
electron glasses~\cite{electron}, and granular materials~\cite{kogranular}.

These systems have in common that, in a part of their phase space, 
they cannot be equilibrated  within the experimental time window:
they undergo a `glass' or `jamming' transition. 
As a result, they exhibit nonequilibrium dynamics, where the
whole history of the sample becomes relevant~\cite{reviewaging}.
The conceptually simplest experiment performed
in glassy materials consists of a sudden quench
of the system into its glassy phase at the initial time $t_w=0$.
The resulting ageing behaviour is best observed through the measurement 
of a two-time quantity, $C(t,t_w)$, typically a 
correlation function or a linear susceptibility.
Experiments show that time translation invariance is absent, i.e.
$C(t,t_w) \neq C(t-t_w)$. Remarkably, 
they also show that the  generic scaling behaviour
$C(t,t_w) \sim {\cal C} (t/t_w)$ is very often observed,
implying that the relaxation time
$t_{\rm rel}$ of the system is of the order 
of its age, $t_{\rm rel} \sim t_w$.

This ageing phenomenology can be explained by models of
coarsening of domains of equivalent thermodynamic 
phases~\cite{reviewaging,reviewalan}.
The canonical example is a ferromagnet quenched into its ordered
phase which exhibits ageing, due to the growth of
ferromagnetic domains of different orientations. A 
length scale $\ell_T(t)$, over which equilibrium is established,
grows with time $t$ at temperature $T$, 
leading to the above scaling behaviour~\cite{reviewalan}.
However, such simple models fail to explain the 
spectacular `memory' and `rejuvenation' effects~\cite{cudean}
observed in more elaborate experimental protocols
unless complemented in some way. 
One approach is to include temperature chaos~\cite{hajime},
which amounts to a great sensitivity of equilibrium states
to temperature changes, as first postulated in the context 
of spin glasses~\cite{chaos}. 
The very existence of temperature chaos in this sense is 
however still debated, with rather negative conclusions.

Ageing may alternatively be viewed as the search by the system 
of the deepest `traps' of a complex energy landscape~\cite{JP}.
Extending this approach to a hierarchical structure of the traps
allows for a qualitative understanding
of temperature cycling experiments~\cite{saclay}.
Multi-trap models of this kind are studied in Ref.~\cite{multitrap}. 
This approach has however the disadvantage of not providing
a real space description of the problem.
Ageing is also captured in a more microscopic way 
by the exact solution of several mean-field disordered 
models~\cite{reviewaging,cuku}.
These do account for rejuvenation and memory~\cite{cuku}, 
but the explanation relies on the notion of
dynamic ultrametricity, which is incompatible 
with experimental data~\cite{cuku}.

In this paper we show that a phenomenological domain growth approach
may be used to interpret $T$-cycling experiments,
{\it provided the dynamics is considered at criticality}.
For a quench into an ordered phase, the equilibrium correlation 
length $\xieq$ is microscopic and therefore decoupled from $\ell_T(t)$.
However, the key point here is that quenching
to a critical point, one has $\xieq = \infty$, with the result 
that equilibrium fluctuations occur on all length scales
up to the dynamic correlation length $\xi(t) = \ell_T(t)$. 
Temperature cycling along a line of critical points thus
affects critical fluctuations of all length scales 
between the lattice spacing and the system size.
This allows for both rejuvenation and
memory effects without making the assumption of temperature chaos,
or evolution in a hierarchical phase space.
This may also be viewed as a concrete realization of the multi-length 
scale scenario discussed by Bouchaud~\cite{JP}.
As an example, the 2D XY model is analytically considered,
although the physical interpretation of the results
is of more general validity.
Finally, we argue that this phenomenological approach may be
relevant to many glassy materials.   

\section{Ageing at a critical point}

The 2D XY model
presents a line of critical points in the temperature
range $T \in [0,T_{\rm KT}]$ where $T_{\rm KT}$ is the
temperature of the Kosterlitz-Thouless transition.
The model describes a two-component order
parameter $\bm{\phi}( \bm{x},t)$ in a bidimensional space.
The properties of the low temperature phase,
$0<T<T_{\rm KT}$, can be described by the spin-wave 
approximation~\cite{ber.71}
\begin{equation}
H [\theta] =  \frac{\rho(T)}{2}  \int \upd^2 \bm{x}
|\bm{\nabla} \theta |^2 ,
\label{spinwave}
\end{equation}
with $\bm{\phi} = e^{i\theta}$ and where $\rho(T)$ is the spin-wave stiffness.
The dynamics is modelled by a Langevin equation
\begin{equation}
\frac{\partial \theta (\bm{x},t)}{\partial t} 
= - \frac{\delta H[\theta]}
{\delta \theta (\bm{x},t)} + \zeta (\bm{x},t).
\label{langevin}
\end{equation}
The last term is the thermal noise, described by
a Gaussian variable of zero mean and variance
$\langle \zeta(\bm{x},t) \zeta(\bm{x'},t') \rangle
= 4 \pi \eta(T) \rho(T) \delta(\bm{x} - \bm{x'})
\delta(t-t')$, where $\eta(T)$ is the usual
critical exponent, linked to the spin-wave stiffness
by $2\pi \eta(T) \rho(T) = T$.
The dynamic behaviour is analysed through the
two-point correlation function
$C(\bm{r},t) = \langle \bm{\phi}(\bm{x},t)
\cdot \bm{\phi}(\bm{x+r},t) \rangle$, and the
two-time autocorrelation function $C(t,t_w) =
\langle \bm{\phi}(\bm{x},t)
\cdot \bm{\phi}(\bm{x},t_w) \rangle$.
Due to the Gaussian character of the Hamiltonian (\ref{spinwave}),
these quantities  follow straightforwardly from
the computation of the angle-angle correlation function
$C_\theta(\bm{k},t,t_w) = \langle \theta (\bm{k},t)
 \theta (- \bm{k},t_w) \rangle$.

The ageing dynamics of the model was discussed in details
in Refs~\cite{cug.94,ber.01,bray}.
In particular, we recall that
topological defects are not described by the spin
wave approximation (\ref{spinwave}), although they can influence the
dynamics following a quench~\cite{ber.01}. 
We thus choose initial conditions such that 
no vortices are initially present in the system,
with $C_\theta(\bm{k},0,0) = 2 \pi \eta(T_i)/k^2$,
where $T_i$ is the initial temperature~\cite{bray}.
This implies $C(\bm{r},0) \sim (r/a)^{-\eta(T_i)}$ at large 
distances. The UV cutoff $a$ is introduced 
through the factor $e^{-k^2 a^2}$ in all integrals
over the Fourier space, simulating the lattice spacing.  
Although the correlation length in the initial state is not 
defined, the correlation function decays very rapidly if 
$\eta(T_i)$ is chosen to have a large value.

The dynamics following a quench to temperature $T_1$
at time $t=0$ is solved by Fourier transforming (\ref{langevin}),
using (\ref{spinwave}).
For the angle-angle correlation one obtains
\begin{equation}
C_\theta(\bm{k},t,t_w) = 
\frac{2 \pi \eta_1}{k^2} \exp[ - (k\ell_{T_1}(t-t_w))^2 ]
+ \frac{2\pi (\eta_i - \eta_1)}{k^2}
\exp [- (k \ell_{T_1}(t+t_w) )^2 ].
\label{ageing1}
\end{equation}
We define the dynamic correlation length $\ell_{T_k}(t) = (\rho_k t)^{1/z}$,
$z=2$, together with the notation $\eta_k = \eta(T_k)$
and $\rho_k = \rho(T_k)$.
Two-point and two-time correlation functions easily 
follow from (\ref{ageing1}). For
times such that $\ell_{T_1}(t-t_w) \gg a$, one finds
\begin{equation}
\begin{aligned}
C(\bm{r},t)  \simeq & \, \exp \left[ - \eta_1 G \left( \frac{r}{2a}
\right) + (\eta_1-\eta_i) G \left( \frac{r}{2\sqrt{2} \ell_{T_1}(t)}
\right) \right], \\
C(t,t_w) \simeq &\left( \frac{a}{\ell_{T_1}(t-t_w)} \right)^{\eta_1}
\left(  \frac{\ell_{T_1}^2(t)+\ell_{T_1}^2(t_w)}{4 \,  \ell_{T_1}(t) \,
 \ell_{T_1}(t_w)}
\right)^{(\eta_1-\eta_i)/2},
\label{ageing}
\end{aligned}
\end{equation}
where $G(x) = \frac{1}{2} \left( C + 2 \ln x + \int_{x^2}^\infty
\upd t \, \frac{e^{-t}}{t}  \right)$; $C$ is the Euler constant.
For distances $a \ll r \ll \ell_{T_1}(t)$, equilibrium
behaviour is obtained at $T_1$, $C(\bm{r}, t) \sim (r/a)^{-\eta_1}$, 
while at large distances, $\ell_{T_1}(t) \ll r$, one has 
$C(\bm{r},t) \sim (r/a)^{-\eta_i} (a/\ell_{T_1}(t))^{\eta_1-\eta_i}$, 
which is the same power law behaviour as in the initial 
state, with a time-dependent multiplicative factor~\cite{bray}.
Similarly, the two-time correlation function
displays an initial equilibrium behaviour $C(t,t_w) \sim 
(t-t_w)^{-\eta_1/2}$, which is interrupted after a time 
$t-t_w \sim t_w$, when the relevant scaling variable
becomes $\ell_{T_1}(t) / \ell_{T_1}(t_w)$.

The dynamics is thus interpreted as consisting of two types of critical 
fluctuations~\cite{ber.01,goodluck}. 
For wavevectors $k \ell_{T_1}(t) \gg 1$, the equilibrium state
at $T_1$ has been reached, while 
fluctuations with $k \ell_{T_1}(t) \ll 1$ are still
very near to their initial nonequilibrium state.
The ubiquitous $t/t_w$ scaling is thus obtained as a consequence
of a growing equilibration length in the system, as in 
ordinary coarsening~\cite{reviewaging}. 
The only difference is that here the correlation length is
fixed to $\ell_T(t)$, while in the former it is microscopic.

\section{Rejuvenation and memory effects}

\begin{figure}
\onefigure[scale=0.6]{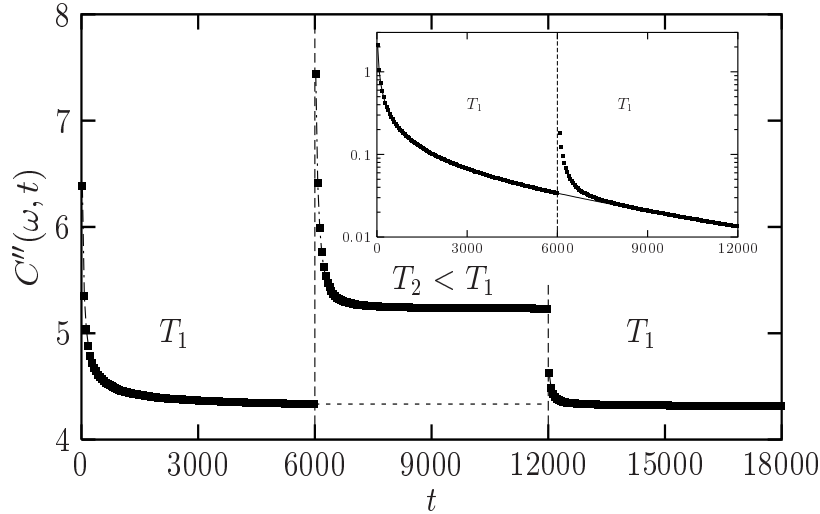}
\caption{Evolution of the imaginary part of the Fourier transform of the 
autocorrelation function in a temperature cycling. 
Parameters are: $a=1$, $\omega=50$, $\eta_i=2.0$, $t_1=6000$, $\eta_1=0.7$,
$\rho_1=1.0$, $t_2=12000$, $\eta_2=0.35$, $\rho_2=0.05$.
Inset: the time interval at $T_2$ has been dropped out. To better
demonstrate the memory, $C''(\omega,t) - C''(\omega,\infty)$ 
has been plotted in a lin-log scale, the full line being the behaviour
with $t_2=t_1$, i.e. without the cycle.}
\label{cycle}
\end{figure}

In temperature cycling experiments, 
the system is first quenched to a temperature $T_1$ where it ages.
After a time $t_1$, the temperature
is shifted to $T_2<T_1$.
It is observed that the ageing is 
restarted by the temperature
change: this is the rejuvenation effect.
Then, at time $t_2 > t_1$, the temperature is changed back to $T_1$.
After a short transient time the ageing is observed to be
the same as if $t_2=t_1$, that is, as if the temperature cycle 
had not existed: this is the memory effect (see Fig.~\ref{cycle}).
The coexistence of these two effects was popularized by the
`dip-experiment' proposed in Ref.~\cite{dip}, which has since been
largely used, showing again very similar results
in different materials~\cite{SG,supra,diel,relax,disferro,saclay,bellon}.

The rejuvenation effect can be conveniently demonstrated 
by taking the limit $t_1 \to \infty$, in which case
the system starts from an equilibrated configuration
at $T_1$ and is quenched to $T_2$.
This is in fact precisely the situation considered in the 
previous section if the changes $T_i \to T_1$ 
and $T_1 \rightarrow T_2$ are made.
Thus, Eqs.~(\ref{ageing}) illustrate
the fact that ageing is restarted when the 
temperature is changed along the line of critical points. 
Keeping $t_1$ finite, the angle-angle correlation becomes,
at times $t>t_w>t_1$,
\begin{equation}
\begin{aligned}
C_\theta(\bm{k},t,t_w) = &
\frac{2\pi \eta_2}{k^2} \exp [-(k \ell_{T_2}(t-t_w))^2 ] 
+ \frac{2\pi (\eta_1 - \eta_2)}{k^2}
\exp [-(k \ell_{T_2}(t+t_w-2t_1))^2 ] \\
+ &  \frac{2\pi (\eta_i - \eta_1)}{k^2}
\exp [-(k \ell_{T_2}(t+t_w-2t_1))^2 ] 
\exp [ -2 (k\ell_{T_1}(t_1))^2 ].
\label{reju}
\end{aligned}
\end{equation} 
When $t_1 \to \infty$, the last term vanishes and 
the results of the previous section are indeed recovered.
On the contrary, when $\ell_{T_2}(t-t_1) \gg \ell_{T_1}(t_1)$, the last two 
terms combine to give an ageing
behaviour as if the system was directly quenched to $T_2$.
Equation (\ref{reju}) shows that fluctuations
with $k \ell_{T_1}(t_1) \gg 1$ age at $T_2$ as if they were
quenched from $T_1$
(the last exponential makes the last term vanish), whereas 
those  with $k \ell_{T_1}(t_1) \ll 1$
age at $T_2$ as in  a quench from $T_i$. In this case the last two terms 
may indeed be combined when the last exponential is $\sim 1$
to make the contribution proportional to $\eta_1$
disappear.

Physically, rejuvenation results from the fact that when 
the temperature is shifted to $T_2<T_1$, all
fluctuations fall out of equilibrium.
Fluctuations such that $k \ell_{T_1}(t_1) \gg  1$ have to adapt their
Boltzmann weight to the new temperature, 
while for $k \ell_{T_1}(t_1) \ll 1$,
no equilibrium had been found at the previous temperature.
Thus {\it even when $t_1 = \infty$, ageing is 
restarted by a temperature change.}
We emphasize that the effect is absent in a 
standard coarsening below $T_c$, where the
reequilibration of thermal fluctuations 
following the shift to $T_2$ is very fast, with an
associated time scale $t_{\rm r}$ given by $\ell_{T_2}(t_{\rm r}) 
\sim \xieq(T_1) \ll \ell_{T_1} (t_1)$.

The memory effect can be seen as the
temperature is changed back to $T_1$ at total
time $t_2$, with $\ell_{T_2}(t_2-t_1) \ll \ell_{T_1}(t_1)$.
The angle-angle correlation function
reads now
\begin{equation}
\begin{aligned}
C_\theta(\bm{k},t,t_w) = &
\frac{2\pi \eta_1}{k^2} 
e^{ -(k \ell_{T_1}(t-t_w) )^2}
+ \frac{2 \pi (\eta_2 - \eta_1)}{k^2}
e^{-(k \ell_{T_1}(t+t_w-2t_2) )^2}
\left[ 1 - e^{-2 (k \ell_{T_2}(t_2-t_1))^2} \right] \\
+ & \frac{2 \pi (\eta_i -\eta_1)}{k^2} 
e^{-2 (k \ell_{T_1}(t_1) )^2} \,  e^{-2 (k \ell_{T_2}(t_2-t_1) )^2} 
e^{-( k \ell_{T_1}(t+t_w-2 t_2))^2}.
\label{memo}
\end{aligned}
\end{equation}
To understand this result, 
three types of fluctuations have to be considered.

(i)
$ k^{-1} \ll \ell_{T_2}(t_2-t_1)$. 
These fluctuations are equilibrated at
$T_2$ and have to reequilibrate at $T_1$.
This can be seen by setting $e^{-2 (k \ell_{T_2}(t_2-t_1))^2} \sim 0$ 
in Eq.~(\ref{memo}), from which one finds an
equation similar to (\ref{ageing1}),
describing a shift from $T_2$ to $T_1$.
The reequilibration takes place on a time scale 
$t_{\rm m}$ such that $\ell_{T_1}(t_{\rm m}) \sim  \ell_{T_2}(t_2-t_1) \ll
\ell_{T_1}(t_1)$, i.e. $t_{\rm m} \ll t_1$.
This is the short initial transient observed 
in experiments and in Fig. 1.

(ii) $\ell_{T_2}(t_2-t_1) \ll k^{-1} \ll \ell_{T_1}(t_1)$. 
Only the first term of Eq.~(\ref{memo}) survives.
These fluctuations
had no time to equilibrate at $T_2$ but were equilibrated
at $T_1$ before the cycle.
Hence, they undergo equilibrium dynamics at $T_1$ 
immediatly after the cycle.
{\it It is in these length scales 
that the memory is conserved.}

(iii)
$\ell_{T_1}(t_1) \ll k^{-1}$. 
Only  the last term in Eq.~(\ref{memo}) 
survives for these wavevectors.
Taking $e^{-2 (k \ell_{T_1}(t_1) )^2} \sim 1$
and $e^{-2 (k \ell_{T_2}(t_2-t_1) )^2} \sim 1$, Eq.~(\ref{ageing1})
is recovered, describing a quench from $T_i$ to $T_1$.
These length scales
retained their initial nonequilibrium state throughout 
the temperature cycle and will thus
equilibrate at $T_1$ through the further growth of $\ell_{T_1}$.

As a consequence, after the transient  time 
$t_{\rm m}$, the dynamics appears as if the cycle at $T_2$
was absent, accounting for the memory effect observed in experiments.
We show in Fig.~\ref{cycle} the behaviour 
of the imaginary part of 
the Fourier transform of the
autocorrelation function 
$C(\omega,t) = \int_0^t \upd t' C(t,t') e^{i \omega (t-t')}$
in a temperature cycle.
At equilibrium, this function is directly 
related by the fluctuation-dissipation
theorem  to the linear susceptibility $\chi(\omega,t)$,
which is the quantity measured experimentally.
The figure demonstrates that the rejuvenation
and memory effects are quantitatively accounted for by the
2D XY model.
The two-time autocorrelation function is computed
from Eqs.~(\ref{reju}) and (\ref{memo}), as was done
to obtain Eq.~(\ref{ageing}) from  Eq.~(\ref{ageing1}). 
As it is rather complicated, we do not report it here, 
its interpretation being the same as that developed above.

Interestingly, the same effects could be observed in a positive 
$T$-cycle, $T_2 > T_1$, provided $\ell_{T_2}(t_2 - t_1) 
\ll \ell_{T_1}(t_1)$. This feature is in qualitative contradiction
with the hierarchical phase space picture~\cite{saclay}, 
and experiments could thus distinguish between the two approaches.

\section{Kovacs effect}

\begin{figure}
\onefigure[scale=0.75]{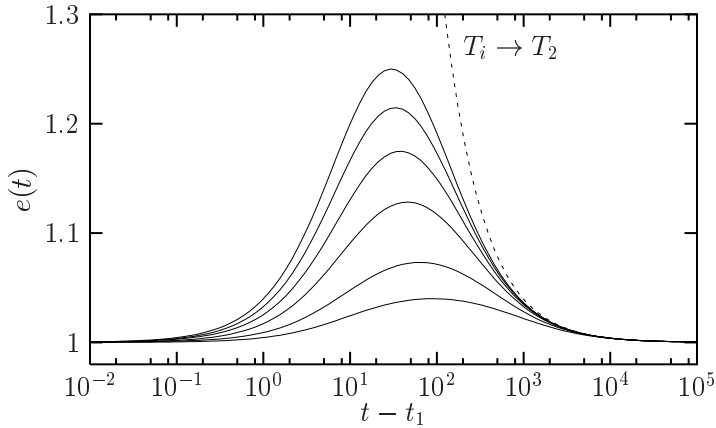}
\caption{Kovacs effect in the 2D XY model. 
Full lines are for various intermediate temperature $T_1=0.95$,
0.9, 0.8, 0.7, 0.6, 0.5 (from bottom to top), with $T_i=5.0$ and 
$T_2=1.0$.}
\label{koko}
\end{figure}

In the literature of structural glasses,  
`memory effect' has a different meaning~\cite{struik,kovacs}. 
Here, the system is first quenched to a given temperature $T_1$.
The slow nonequilibrium evolution of a one-time quantity $V(t)$
(volume, refraction index) is followed.
The temperature is then raised at time $t_1$ to a value $T_2$
such that, at equilibrium, $V$  would have 
the value it has just before
the shift, i.e. $V(t_1) = V_{\rm eq}(T_2)$.
Whereas it would be naively expected that
$V(t>t_1) = const$, a nontrivial non-monotonic  behaviour 
is observed instead. To avoid confusion, we will refer 
to this effect as the Kovacs effect~\cite{kovacs}. 
Again, we note that it was observed in several 
materials~\cite{struik,diel,kogranular,kovacs,kofoam,glycerol,koea}.

Within the 2D XY model, one follows the behaviour 
of the energy density $e(t)$ in a protocol
$T_i \to T_1$ at $t=0$ and $T_1 \to T_2$ at $t_1$, 
with $T_i > T_2 > T_1$. For times such that $\ell_{T_2}(t-t_1)\gg a$, we get
in the harmonic limit
\begin{equation}
e(t) = \frac{1}{16\pi} \left( \frac{2T_2}{a^2} +
\frac{T_i - T_1}{\ell_{T_2}^2(t-t_1) + 
\ell_{T_1}^2(t_1)} + \frac{T_1-T_2}
{\ell_{T_2}^2(t-t_1)}  \right).
\label{kov}
\end{equation}
The energy first increases due to the third term, since 
$\eta_1 < \eta_2$. When $\ell_{T_2} (t-t_1) \gg \ell_{T_1}(t_1)$,
the second and third terms combine to give 
the simple ageing behaviour of a system directly 
quenched from $T_i$ to $T_2$, and hence a decreasing energy density 
which merges with  the curve obtained in this simple experiment.
Equation (\ref{kov}) was used to build the curves in Fig.~\ref{koko},
which reproduce quantitatively experimental findings.

Immediately after the shift at time $t_1$, fluctuations with
$k^{-1} \ll  \ell_{T_1}(t_1)$ are equilibrated at $T_1$ and have to increase
their energy in order to equilibrate at the new temperature $T_2$.
Fluctuations with $\ell_{T_1}(t_1) \ll k^{-1}$, which retained their
nonequilibrium state at $T_i$,  decrease their energy
to reach their equilibrium state at $T_2$.
These two types of fluctuations have opposite 
contributions to the energy evolution and act on different 
time scales.
Thus, a maximum occurs at a time $t_{\rm k}$, such that
$\ell_{T_2}(t_{\rm k}-t_1) \sim \ell_{T_1}(t_1)$,
the larger $T_2-T_1$ the larger the height of the maximum.
This interpretation can be viewed as a spatial 
transposition of the distribution of time scales usually 
invoked to account for this effect~\cite{struik}.

\section{Discussion}
We have shown that coarsening models at criticality
reproduce generic ageing, rejuvenation,
memory and Kovacs effects seen in many glassy systems,
thus providing a new and simple phenomenological
domain growth approach to ageing phenomena.

This seems to beg the question: are
glassy systems critical? 
A possible answer is yes they are, in which case
glassy systems should have in common a line of critical points
below the glass transition, in analogy with the 2D XY model. 
This is not inconsistent with existing theoretical descriptions
of the spin glass phase~\cite{reviewaging}, but seems less likely
in other systems, such as disordered ferromagnets~\cite{disferro} or 
ferroelectrics~\cite{diel}.
Another is no; but they do at least behave as if there is a line of
critical points
over the experimental time window. The second, more pragmatic
solution requires that the coherence length $\ell_T(t)$ is
never decoupled from $\xieq(T)$, even when the latter is finite.
One might argue this if the excitations on length scales
exceeding $\xieq(T)$ are dominated by activated processes, 
giving exponentially increasing time scales for $\ell_T(t) \ge \xieq$.
These conditions would make it virtually impossible to enter
a regime where $\ell_T(t) \gg \xieq(T)$. In this case, a slow evolution
in the crossover region could result in effective 
critical behaviour with continuous evolution of the exponents,
with the temperature, which is the main ingredient of our analysis of
the rejuvenation effect.
This is fully consistent with results in 3D spin glasses~\cite{koea,JP2}.
It also implies that similar results could be
obtained in spin glass films around the $T=0$ critical point~\cite{film}.

In both cases, of course, disorder is essential.
Indeed, we would not expect to observe ageing phenomena
easily in a non-disordered critical system as
the divergence of time scales is actually very weak
and can in general be beaten easily, both experimentally and numerically.
Disorder would therefore be needed to provide the necessary separation of
time scales~\cite{JP} and to
put the ageing phenomena discussed here
within the observed experimental time window.

\end{document}